\documentclass[11pt,a4paper]{article}
\usepackage{jheppub}
\usepackage[latin9]{inputenc}
\usepackage{amsmath}
\usepackage{amssymb}
\usepackage{setspace}
\usepackage{graphicx}
\usepackage{babel}

\usepackage{physics}
\usepackage{bm}

\begin{document}
\title{
Phase Boundary of Nuclear Matter in Magnetic Field}

\author[a]{Yuki Amari,
}
\emailAdd{amari.yuki@keio.jp}

\author[a,b]{Muneto Nitta
}
\emailAdd{nitta@phys-h.keio.ac.jp}
\author[a]{and Zebin Qiu
}
\emailAdd{qiuzebin@keio.jp}

\affiliation[a]{%
Department of Physics $\&$ Research and Education Center for Natural Sciences, Keio University, 4-1-1 Hiyoshi, Yokohama, Kanagawa 223-8521, Japan}
\affiliation[b]{
International Institute for Sustainability with Knotted Chiral Meta Matter(WPI-SKCM$^2$), Hiroshima University, 1-3-2 Kagamiyama, Higashi-Hiroshima, Hiroshima 739-8511, Japan
}

\abstract{
Nuclear matter with a strong magnetic field is prevalent inside neutron stars and heavy-ion collisions. 
In a sufficiently large magnetic field 
the ground state is either a chiral soliton lattice (CSL), an array of solitons of the neutral pion field, or a domain-wall Skyrmion phase in which Skyrmions emerge inside the chiral solitons. In the region of large chemical potential and a magnetic field lower than its critical value for CSL, a Skyrmion crystal is expected to take up the ground state based on the chiral perturbation theory at the next leading order.
We determine the phase boundary between such a Skyrmion crystal and the QCD vacuum. 
There was a conjecture that a magnetic field deforms the Skyrmion into a pancake shape 
whose boundary is a superconducting ring of charged pions. 
In contrast, through the exact Skyrmion solution, we find that the pancake conjecture holds approximately in
a strong magnetic field, but fails for a weak one.
We also validate that a Skyrmion would shrink to null without the Skyrme term, although Derrick's scaling law is modified by a background magnetic field, and the stability at the leading order is not ruled out in theory.
}

\maketitle

\section{Introduction: QCD Phases in a Magnetic Field}

It is established that baryons and mesons are composite particles made of quarks glued by gluons in quantum chromodynamics (QCD), the fundamental theory of strong interaction.
At low energy, 
QCD is effectively described by the chiral perturbation theory (ChPT) constructed in terms of light degrees of freedom,  
which are approximated by Nambu-Goldstone bosons (NGBs) 
associated with spontaneous chiral symmetry breaking.
The lightest relevant NGBs are the pions subject to $SU(N_f=2)$ symmetry in the 2-flavor case.
In such a theory, baryons can be described 
as solitonic objects, namely the Skyrmions~\cite{Skyrme:1961vq,Skyrme:1962vh} 
supported by the third homotopy group 
$\pi_3 (S^3) \simeq {\mathbb Z}$. 
The Skyrme model is 
justified at large $N_c$ (color number) limit to characterize nucleons properly 
\cite{Witten:1983tw,Adkins:1983ya,Adkins:1983hy,Zahed:1986qz}. 
Then a Skyrmion is naturally generalized to a Skyrmion crystal for studying dense baryonic matter~\cite{Klebanov:1985qi,Goldhaber:1987pb,Kugler:1988mu,Kugler:1989uc,Castillejo:1989hq}.   
Skyrmion crystals prove applicable to dense matter such as compact stars, bringing the Skyrme model and its various extensions back to the attention of the rising crossover research in nuclear astrophysics, among others.
See, e.g., refs~\cite{SilvaLobo:2009qw, Nelmes:2012uf, Canfora:2013xja, Liu:2018wgv, Yang:2020ucv, Lee:2021hrw, Adam:2022aes, Ma:2023ugl, Rho:2023vvs, Leask:2023tti} 
for recent studies and 
refs.~\cite{Ma:2019ery,Adam:2023cee,Leask:2024hcu} 
for recent reviews.

On the other hand, QCD phase diagram under extreme conditions, such as temperature, density and external fields, has been a longstanding hot issue.
Particular to our interest is the magnetic field given its relevance in neutron stars and heavy-ion collision experiments. 
In a magnetic field, there are a couple of different ways for baryons to emerge with topological charges.
The first homotopy group features the 
chiral soliton lattices (CSL) 
that is an array of solitons made of the neutral pion $\pi^0$ 
\cite{Son:2007ny,
Eto:2012qd,
Brauner:2016pko,Brauner:2017uiu,Brauner:2017mui,Brauner:2023ort} 
or $\eta$ (or $\eta^\prime$) meson \cite{Huang:2017pqe, Nishimura:2020odq,Chen:2021aiq,
Eto:2021gyy,Eto:2023rzd}, 
or their mixture as a quasicrystal \cite{Qiu:2023guy}.\footnote{ 
CSLs were also proposed in 
QCD-like theory such as 
$SU(2)$ QCD, vector-like gauge theories \cite{Brauner:2019rjg,Brauner:2019aid} and  supersymmetric QCD~\cite{Nitta:2024xcu}.
}
Thermal fluctuations enhance their 
stability~\cite{Brauner:2017uiu,Brauner:2017mui,Brauner:2021sci,Brauner:2023ort}.
In the common magnetic field, captured by the third homotopy group are
domain-wall (DW) Skyrmions \cite{Eto:2023lyo,Eto:2023wul,Eto:2023tuu,Amari:2024mip,Amari:2024fbo},
vortex Skyrmions \cite{Qiu:2024zpg}, 
(magnetized) Skyrmion crystals
\cite{Kawaguchi:2018fpi,Chen:2021vou, Evans:2022hwr,Evans:2023hms} 
and baryonic tubes \cite{Canfora:2018rdz,Canfora:2020kyj,
Cacciatori:2021neu}, among others.

The Wess-Zumino-Witten (WZW) term matching anomaly at low energy 
\cite{Son:2004tq,Son:2007ny}
plays a crucial role to realize these states as ground states.
Particularly worth reviewing is the $\pi^0$ CSL, the lightest mesonic soliton configuration that arises in a baryon-rich context. It takes up the ground state under the condition of an external magnetic field 
larger than the critical value 
\cite{Son:2007ny,Brauner:2016pko} 
\begin{align}
    B \geq B_{\rm CSL} \equiv \frac{16 \pi m_\pi f_\pi^2}{e \mu_{\rm B}}.  \label{eq:B-CSL}
\end{align}
Among it, $f_\pi$ is the pion decay constant and $m_\pi$ is the effective pion mass.
As a consequence of the WZW term, a single chiral soliton carries the baryon number $e B/2\pi$ per unit surface area. 
And if we cut a surface of the area $2\pi/eB$ in the CSL it contains a fermion \cite{Amari:2024fbo}.
Moreover, within the CSL phase 
at higher density/larger magnetic field 
,
Skyrmions appear on top of the lattice to form a composite state, 
called DW Skyrmions 
\cite{Nitta:2012wi,Nitta:2012rq,Gudnason:2014nba,
Gudnason:2014hsa}, with the corresponding ground state referred to as DW Skyrmion (DWSk) phase 
\cite{Eto:2023lyo,Eto:2023wul,Eto:2023tuu,Amari:2024mip,Amari:2024fbo}.
Intriguingly, a single DW Skyrmion carries the baryon number two and has the nature of a boson \cite{Amari:2024fbo}.

We now turn to the region of higher density and a magnetic field lower than the critical one of CSL 
. 
In such a region, 
nuclear matter as a Skyrmion crystal 
is anticipated via the Skyrme model, at least at zero magnetic field, as mentioned above
\cite{Klebanov:1985qi}. 
Our interests here are how it is altered by a finite magnetic field and its relation with the CSL.
Along this line, 
some studies were made before \cite{Kawaguchi:2018fpi,Chen:2021vou}. 
In order to advance the understanding, 
let us recall the argument of 
Son and Stephanov that a Skyrmion can be regarded as 
a chiral soliton with a finite size and in the shape of 
a pancake (or a disk), whose boundary encompasses charged pions with their phase winding as a superconducting vortex
\cite{Son:2007ny}. 
From the quantization of such a conjectured charged pion 
vortex, the pancake is supposed to have the transverse area prescribed by the superconducting ring
\cite{Son:2007ny,Amari:2024mip}:  
\begin{align}
    S_{N_{\rm B}} = \frac{2\pi N_{\rm B}}{e B}, 
    \quad 
    N_{\rm B} \in {\mathbb Z} 
    \label{eq:quantization}
\end{align}
where $N_{\rm B}$ means the baryon number. 
This equation correctly reproduces the baryon density of CSL, i.e., 
$N_{\rm B}/S_{N_{\rm B}} = e B/2\pi$. 
On the other hand, by using 
the tension 
$T_{\rm CS} = 8 m_{\pi} f_\pi^2$
of a single chiral soliton (without the contribution from the WZW term, which would soon be addressed),
the energy of a single Skyrmion 
($N_{\rm B}=1$) configured as a pancake of the surface area 
in eq.~(\ref{eq:quantization})
can be evaluated as
\begin{align}
    E_{\rm pan} = T_{\rm CS} S_1 =
    \frac{16 \pi m_\pi f_\pi^2}{e B}.
\end{align}
Then, the critical chemical potential 
for the pancake Skyrmion to be created would require
\begin{align}
   E_{\rm pan}^{\text{tot}} = E_{\rm pan} 
   - \mu_{\rm B} < 0
    \quad \leftrightarrow \quad 
    B \geq B_{\rm pan} \equiv
    \frac{16 \pi m_\pi f_\pi^2}{e \mu_{\rm B}} 
\end{align}
where the $E_{\rm pan}^{\text{tot}}$ is the total energy featuring the contribution from the WZW term $-\mu_{\rm B}N_\text{B}$.
We emphasize that such critical $B_\text{pan}\left(\mu_\text{B}\right)$ is meaningful only 
in the sense of the phase boundary between vacuum and Skyrmion crystal (of pancake Skyrmions) because a phase transition into a certain Skyrmionic phase always begins with the nucleation of a single Skyrmion. 
As one can see,  
$B_\text{pan}$ coincides
with that of the CSL in eq.~(\ref{eq:B-CSL}). 
Does this imply that there is no Skyrmion crystal phase outside the CSL phase? 
Instead we suspect it could be an artifact brought by approximating the Skyrmion shape as a pancake.
Thus, in order to determine the phase boundary more precisely, 
we should explicitly construct 
Skyrmion solutions in the magnetic field. 
The phase boundary, if exists, 
should end at $\mu_{\rm B}=m_N$, the nucleon mass at zero magnetic field $B=0$. 
Generally the $\mu_{\rm B}$ for the phase boundary is a function of $B$, which indicates the $B$-dependence of the nucleon mass.
Then, does such curve $\mu_\text{B}(B)$ cross the CSL phase boundary at large magnetic field? 
These are the questions that will be answered by the present paper. 

Another related investigation of this paper is on scaling properties of Skyrmions in the magnetic field. 
The purpose is twofold.
One is to compare our exact solution with the pancake Skyrmion conjectured by Son and Stephanov~\cite{Son:2007ny}.
Among the Skyrmion profile, we observe a ring on which the magnitude of charged pions is maximized and calculate the radius of the ring with dependence on the magnetic field.
We conclude that the Son and Stephonov's quantization condition in eq.~(\ref{eq:quantization}) is almost satisfied for the aforementioned ring area at a larger magnetic field, which means the pancake conjecture is a fairly good approximation therein. 
However as we will show, the ring radius deviates from that given by eq.~(\ref{eq:quantization}) when the magnetic field is small. 
The other motivation to study typical scales of a Skyrmion is to see whether it can be stabilized by a background magnetic field at the leading order of ChPT, i.e., ${\cal O}(p^2)$,
without the Skyrme term which is of the next leading order ${\cal O}(p^4)$.
As is well known, if there is no gauge field, such ${\cal O}(p^4)$
terms in kinetic energy are needed to evade the Derrick scaling law~\cite{Derrick:1964ww}, yielding a finite size of Skyrmion. 
Otherwise, the solution is unstable in the sense that the Skyrmion tends to shrink to an infinitesimal space.
The idea of leveraging a gauge field is inspired by the finding that whereas gauged $O(3)$ lumps in 2+1 dimensions are unstable without a potential term, they can be stable in a background magnetic field \cite{Amari:2024adu}. 
Moreover, DW Skyrmions are stable without the Skyrme term at strong gauge coupling, and DW anti-Skyrmions are stable in the whole range of parameters without a Skyrme term \cite{Amari:2024fbo}.
Even more relevant is the discovery in Ref.~\cite{Gudnason:2024ers} that an external electric field could prevent Skyrmions from shrinking when the Skyrmion charge is taken into account properly. 
We thus scrutinize the stability of a Skyrmion at 
${\cal O}(p^2)$ without the Skyrme term 
in the presence of a background gauge field, aiming at (Skyrme) model independent conclusions. 
Indeed, we show that Derrick's scaling law supports this possibility
(in Appendix~\ref{sec:Derrick}).
However, by direct numerical computations,
we negate the possibility with a conclusion that 
Skyrmion cannot be stable at ${\cal O}(p^2)$ 
even in the presence of a background gauge field.

This paper is organized as follows.
In Sec.~\ref{sec:CSL-DWSk} we review CSL and DWSk phases.
In Sec.~\ref{sec:SkX} we explore the phase boundary of  
Skyrmion crystal against QCD vacuum, and its intersection with that of CSL. 
In Sec.~\ref{sec:op2} we study scaling properties of the Skyrmion in a magnetic field, analyzing the pancake Skyrmion conjecture and a Skyrmion without the Skyrme term.
Sec.~\ref{sec:summary} is devoted to a conclusion.
In Appendix \ref{sec:Derrick}
we present Derrick's scaling law modified by a background gauge field. 

\section{Review: Chiral Soliton Lattice and Domain-Wall Skyrmion}\label{sec:CSL-DWSk}

In this section we review two known low-energy QCD phases under a magnetic field, to be compared with the Skyrmion crystal, which describes nuclear, or generally put, baryonic matter. 
The physics regime that we discuss is governed by chiral symmetry breaking
and quark condensate. Under such assumptions, we adopt the ChPT, a
momentum expansion subject to a power counting of $\mathcal{O}\left(p^{n}\right)$.
The leading order Chiral Lagrangian includes kinetic and mass terms:
\begin{equation}
\mathcal{L}_{\text{ChPT}}=-\frac{f_{\pi}^{2}}{4}\mathrm{Tr}\left(L_{\mu}L^{\mu}\right)+\frac{f_{\pi}^{2}m_{\pi}^{2}}{2}\mathrm{Tr}\left(\Sigma-1\right).\label{eq:Lkin}
\end{equation}
In the present study, we deal with two-flavor $\Sigma=\exp\left(i\boldsymbol{\tau}\cdot\boldsymbol{\varphi}\right)\in SU\left(N_{f}=2\right)$ 
for which we define the covariant left-handed and right-handed currents
\begin{equation}
L_{\mu}=\Sigma^{\dagger}D_{\mu}\Sigma,\quad R_{\mu}=\Sigma D_{\mu}\Sigma^{\dagger},
\label{eq:LR}
\end{equation}
respectively, with covariant derivative encompassing the $U(1)$
electromagnetic gauge field $A_{\mu}$, i.e.,
\begin{equation}
D_{\mu}\Sigma=\partial_{\mu}\Sigma-i e A_{\mu}\left[Q,\Sigma\right];\quad Q=1/6+\tau^{3}/2.
\end{equation}
We restrict our discussions to a homogeneous external axial magnetic field
set along the longitudinal axis $\boldsymbol{B}=B\hat{z}$. 
Then, an azimuthal $A_{\phi}=Br\sin\theta/2$
(spanned on spherical coordinates) suffices for the scenario.

To our interest are QCD phases at finite $B$ and baryon density $\mu_{\rm B}$.
The latter is captured effectively by a baryon gauge field 
$A_{\mu}^{\text{B}}=\left(\mu_{\rm B},\boldsymbol{0}\right)$. Thereafter,
effects from the triangle anomaly should be taken into account. In the $SU(2)$
ChPT framework, the anomaly is encoded in the WZW term
\begin{equation}
\mathcal{L}_{\text{WZW}}=J_{\text{B}}^{\mu}\left(qA_{\mu}+A_{\mu}^{\text{B}}\right),
\end{equation}
with $q=e/2$ and the topological Goldstone-Wilczek current \cite{Goldstone:1981kk,Witten:1983tw} interpreted as the baryon current in our context 
\begin{equation}
J_{\text{B}}=\frac{1}{24\pi^{2}}\mathrm{Tr}\left\{ l\wedge l\wedge l-3i e Qd\left[A\wedge\left(l-r\right)\right]\right\} ,\label{eq:JB}
\end{equation}
for which we employ $l=\Sigma^{\dagger}d\Sigma$ and $r=\Sigma d\Sigma^{\dagger}$
to write things tersely with differential forms. One can refer to ref.~\cite{Son:2007ny} for
an alternative expression of $J_{\text{B}}$ using covariant forms eq.~\eqref{eq:LR}. Configurations with finite baryon number $N_{\text{B}}=\int d^{3}xJ_{\text{B}}^{0}$
indicate a baryonic phase with $N_{\text{\text{B}}}$ conserved as a topological charge.

\begin{figure}      
\centering       
\includegraphics[width=0.75\columnwidth]{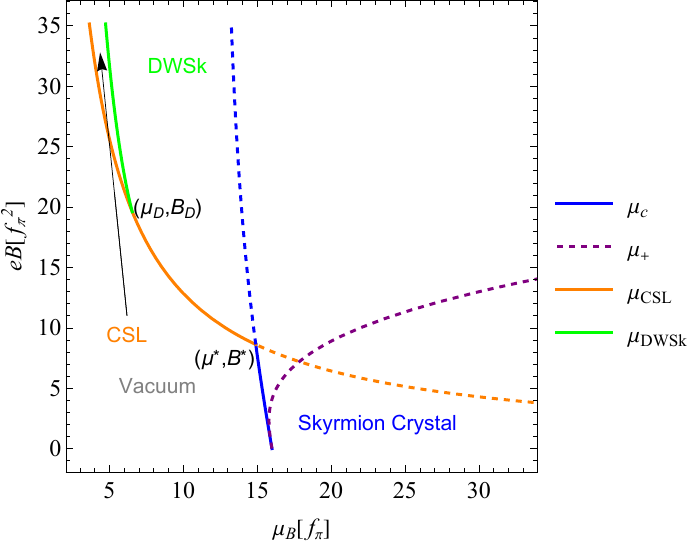}      
\caption{Phase diagram of low energy dense QCD in terms of the baryon chemical potential $\mu_{\rm B}$ and the external magnetic field $B$. 
Curves represent critical baryon chemical potentials as functions of the magnetic field (or vice versa), above the values of which the corresponding state could be a ground state. $\mu_c$ is for a Skyrmion and $\mu_+$ is for an excited Skyrmion (spin anti-parallel to the magnetic field). They are our original results depending on the choice of the Skyrme parameter $s$ (among Eq.~\eqref{eq:Lskyr}), as detailed in Sec.~\ref{sec:SkX}. $\mu_\text{CSL}$ for CSL and $\mu_\text{DWSk}$ for DW Skyrmion are model independent, as reviewed in Sec.~\ref{sec:CSL-DWSk}.
}    
\label{fig:phase}  
\end{figure}

It is known from ref.~\cite{Brauner:2016pko} that when $B$ is strong,
the charged degrees of freedom $\varphi_{1,2}$ decouple from the
dynamics in terms of the ground state, leaving the neutral pion $\pi^{0}$, i.e., $\Sigma\rightarrow\exp\left(i\tau_3\varphi_3\right)$,
to form a soliton domain wall, a.k.a CSL. Technically this can be
seen from the Lagrangian after dimensional reduction
\begin{equation}
\mathcal{L}_{\text{ChPT}}+\mathcal{L}_{\text{WZW}}\rightarrow\frac{f_{\pi}^{2}}{2}\left(\frac{\partial\varphi_3}{\partial z}\right)^{2}-f_{\pi}^{2}m_{\pi}^{2}\left(1-\cos\varphi_3\right)+\frac{e\mu_{\rm B} B}{4\pi^{2}}\frac{\partial\varphi_3}{\partial z}.\label{eq:LCSL}
\end{equation}
In this study we adopt constant $f_\pi=54 \text{ MeV}$ and $m_\pi=138 \text{ MeV}$, taken from ref.~\cite{Adkins:1983hy}.
\footnote{This seemingly unconventional choice of $f_\pi$ is designed to fit the experimental data of nucleon mass in the Skyrme model. However, physics presented in this section is (Skryme) model independent and holds for general values of $f_\pi$ and $m_\pi$.}
To have the lowest energy, $\pi^0$
exhibits homogeneous transverse (the $xy$-plane) distribution. The action principle with $\partial_{t,x,y}\varphi_3=0$ leads
to a sine-Gordon soliton with longitudinal (the $z$-axis) periodicity.
The elliptic modulus and related physical quantities such as energy
or period are determined by $\mu_{\rm{B}} B$. Importantly, for a given
$B$, there exists a critical density (baryon chemical potential)
\begin{equation}
\mu_{\text{CSL}}=16\pi f_{\pi}^{2}m_{\pi}/e B,\label{eq:muCSL}
\end{equation}
representing the same physics as eq.~(\ref{eq:B-CSL}), as plotted in the orange (solid annexed with dotted, whose connotation would be detailed later) curve in fig.~\ref{fig:phase}. 
Only when $\mu_{\rm{B}}>\mu_{\text{CSL}}$ can the CSL arise from QCD vacuum. 
The physical mechanism of this critical phenomenon consists in the baryon number carried by CSL, i.e.,
$ N_{\rm{B}}^{\text{CSL}}=eB\int{d^3x \partial_z\varphi_3/{4\pi^2}} $
receiving a contribution of $e BS/2\pi$
from each period where $\varphi_3$
varies $2\pi$ and occupies transverse area $S$. 
The $N_{\rm{B}}^{\text{CSL}}$ per soliton equals the $N_{\rm{B}}$ in eq.~(\ref{eq:quantization}) 
per an isolated Skyrmion if the Skyrmion is in the shape of a pancake with thickness equal to the CSL period and transverse area identical to $S$, albeit the two configurations have different underlying homotopy.
As long as $N_{\text{B}}\neq0$, the
WZW term would decrease the free energy by $-\mu_{\rm{B}} N_{\text{B}}$ as
seen from the last term on the r.h.s of eq.~\eqref{eq:LCSL}. 
Therefore, we expect a similar critical phenomenon for a single Skyrmion to arise from the vacuum, which stands for the boundary between the nuclear phase and the vacuum. It is our main motivation to draw such a curve, the critical baryon chemical potential $\mu_c\left(B\right)$ on the phase diagram, compared with eq.~\eqref{eq:muCSL}.

In the phase diagram, we intend to add one more relevant phase; the DWSk phase discovered in ref.~\cite{Eto:2023lyo,Eto:2023wul,Eto:2023tuu,Amari:2024mip,Amari:2024fbo}.
The DW Skyrmions essentially result from the addition of $\pi^{\pm}$
winding to the $\pi^{0}$ domain wall so the homotopy is equivalent
to Skyrmion's $\pi_{3}\left(S^{3}\right)$. The difference is, DW
Skyrmions emerge on top of CSL background so such a phase always resides within the CSL realm. The critical chemical potential of 
the DWSk phase $\mu_{\text{DWSk}}$ can be determined analytically 
\begin{equation}
\mu_{\text{DWSk}}=\frac{16\pi f_{\pi}^{2}}{3m_{\pi}}\kappa^{-3}\left[\left(2-\kappa^{2}\right)E\left(\kappa\right)-2\left(1-\kappa^{2}\right)K\left(\kappa\right)\right],\label{eq:muDW}
\end{equation}
with the complete elliptic integrals of the first kind $K\left(\kappa\right)$
and of the second kind $E\left(\kappa\right)$. 
Here $\kappa$ ($0 \le \kappa \le 1$) is the elliptical
modulus of the sine-Gordon  soliton $\varphi_3$ solved from eq.~\eqref{eq:LCSL},
which is an implicit function of $\mu_{\rm{B}} B$ determined by minimizing
the CSL energy, as detailed in refs.~\cite{Eto:2023wul,Amari:2024fbo}. One can prove $\mu_{\text{DWSk}}\left(\kappa\rightarrow1\right)=\mu_{\text{CSL}}$,
and the phase boundary between DWSk and CSL phases ends on 
\begin{eqnarray}
    (\mu_D,B_D) 
    &=& 
    \left(\frac{16\pi f_{\pi}^2}{3m_{\pi}}, 
    \frac{3m_{\pi}^2}{e} \right), \label{eq:TCP}
\end{eqnarray}
which shows in a technical sense that the DWSk phase is bounded within the CSL phase.
The relation between $\mu_{\text{DWSk}}$ and $B$ is plotted as the green curve in fig.~\ref{fig:phase}. It is worth clarifying that analysis on CSL and DWSk phases  
is model-independent. 
Also, we remark that $\mu_D$ evaluated here proves almost only $1/3$ of that estimated in ref.~\cite{Eto:2023lyo}. The reason is the discrepancy in the choice of $f_\pi$.


\section{Skyrmion Crystal Phase Boundary}
\label{sec:SkX}

Now we tackle the main issue: at finite
$\mu_{\rm B}$ and $B$, what is the critical $\mu_{\rm B} $ for 
the Skyrmion crystal phase
to be the ground state, competing with $\mu_{\text{CSL}}$.
In the Skyrmion description of baryons, the spatial integration of
$J_{\text{B}}^{0}$ from eq.~\eqref{eq:JB} proves an integer $N_{\text{B}}$
independent of $A_{\mu}$. It originates from the $l\wedge l\wedge l$
term, which vanishes in the CSL case. Such essential difference is epitomized
by the distinction bewteen $\pi_{3}(S^{3})$ of Skyrmion
and $\pi_{1}(S^{1})$ of CSL. 
To draw their comparison
on the phase diagram, we focus on the $N_{\text{B}}=1$ case because the nucleation starts from a single Skyrmion. To have a stable Skyrmion solution
evading the Derrick's scaling law, we further incorporate the Skyrme
term in the Lagrangian:
\begin{equation}
\mathcal{L}_{\text{Skyr}}=\frac{1}{32s^{2}}\mathrm{Tr}\left[L_{\mu},L_{\nu}\right]\left[L^{\mu},L^{\nu}\right].\label{eq:Lskyr}
\end{equation}
We would address the nuanced scaling law modified by $B$ in the subsequent Sec.~\ref{sec:op2}, where the dimensionless Skyrme parameter $s$ is varied to explore
the model dependent (or independent) nature of our conclusion. 
For now, we would take the fixed value $s=4.84$, coming from fitting the experimental data of the nucleon mass~\cite{Adkins:1983hy}.
Of course such fitting already means the phase boundary ends at 
the $\mu_c\left(B=0\right)=m_N$. 
The question is how such a phase boundary extends for $B \neq 0$.

In our preceding work ~\cite{Chen:2023jbq}, the gauged Skyrme model with Lagrangian $\mathcal{L}_{\text{ChPT}}+\mathcal{L}_{\text{Skyr}}$
had been solved for the same scenario of $A_{\mu}$ but $\mu_{\rm B}$ was
NOT included. The present work further incorporates a finite $\mu_{\rm B}$
among the $\mathcal{O}\left(p^{4}\right)$ Lagrangian 
\begin{equation}
\mathcal{L}^{\left(4\right)}=\mathcal{L}_{\text{ChPT}}+\mathcal{L}_{\text{Skyr}}+\mathcal{L}_{\text{WZW}},
\label{eq:L4}
\end{equation}
which is quintessential for the phase diagram. The topological WZW
term contributes to the energy a constant shift $\propto \mu_{\text{B}}B$. 
Thus, it would not alter the equation of
motion (EOM) or the solution profile of ref.~\cite{Chen:2023jbq}. We therefore follow the transcript therein to evaluate the Skyrmion mass $M=\int{d^3xT^{00}}$ from the energy momentum tensor $T^{\mu\nu}$ corresponding to $\mathcal{L}^{\left(4\right)}-\mathcal{L}_{\text{WZW}}$.

\begin{figure}      
\centering       
\includegraphics[width=0.75\columnwidth]{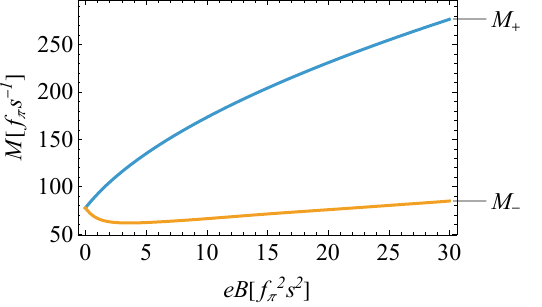}      
\caption{Skyrmion masses of the lowest (internal) energy state $M_{-}$ and excitated state $M_{+}$.}     
\label{fig:Mpm}  
\end{figure}

For completeness, we point out 
there are two independent solutions for $M\left(B\right)$, up to the relative orientation of Skyrmion magnetic moment towards $\boldsymbol{B}$. ref.~\cite{Chen:2023jbq} presents only one branch among the two, which is of the lower energy, so let us denote that as $M_{-}\left(B\right)$. The other branch with $M_{+}\left(B\right)$ is worth mentioning because the physical interpretation of such bifurcation of $M$ is related to Skyrmion versus anti-Skyrmion. Especially, in the present work at finite $\mu_\text{B}$, anti-Skyrmion would feature a $\mathcal{L}_{\text{WZW}}$ with a flipped sign compared to that of Skyrmion, which is indeed relevant in eq.~\eqref{eq:L4}. 
To further explain this point, let us take a closer look at the axial hedgehog Ansatz tailored based on the symmetries of the present scenario:
\begin{equation}
\varphi_{1}+i\varphi_{2}=f\sin g\exp\left( i \epsilon_{1} \phi\right),\quad\varphi_{3}= \epsilon_{2} f\cos g.\label{eq:ansatz}
\end{equation}
Here the $\phi$ is the azimuthal angle given the cylindrical symmetry.
$f$ and $g$ are functions of $r$ and $\theta$, which will be solved
from the EOM governed by the variational principle $\delta M=0$.
Generally, there is freedom to choose the signs $\epsilon_{1,2}=\pm1$.
The choice of $\epsilon_{1}$ leads to the bifurcated
$M_{\pm}$ whose integrand $T_{\pm}^{00}$ reads:
\begin{align}
T_{\pm}^{00}= & \frac{f_{\pi}^{2}}{2}\left(\left|\nabla f\right|^{2}+\sin^{2}f\left|\nabla g\right|^{2}+\sin^{2}f\sin^{2}g\varUpsilon_{\pm}^{2}\right)\nonumber \\
 & +\frac{1}{2s^{2}}\sin^{2}f\left[\left|\nabla f\times\nabla g\right|^{2}+\varUpsilon_{\pm}^{2}\left(\left|\nabla f\right|^{2}+\sin^{2}f\left|\nabla g\right|^{2}\right)\right]\nonumber \\
 & +f_{\pi}^{2}m_{\pi}^{2}\left(1-\cos f\right),
 \label{eq:T00}
\end{align}
in which we defined 
\begin{equation}
\Upsilon_{\pm}=\frac{1}{r\sin\theta}\pm\frac{1}{2}Br\sin\theta.
\label{eq:upsilon}
\end{equation}
We solved the two cases 
with the common boundary conditions
\begin{equation}
f(r=0,\,\theta)=\pi,\quad f(r=\infty,\,\theta)=0,\quad g(r,\,\theta=0)=0,\quad g(r,\,\theta=\pi)=\pi,
\end{equation}
and present the resulting $M_{\pm}$ in fig.~\ref{fig:Mpm}.
Obviously we observe $M_{+}>M_{-}$ so the configuration with $\epsilon_1=+1$ in eq.~\eqref{eq:ansatz}
proves irrelevant to the discussion of ground state. However, it does
have physical meaning as an excitation state, in view that its magnetic
moment 
\begin{equation}
\boldsymbol{m}\equiv\boldsymbol{x}\times\boldsymbol{j}_{Q};\quad j_{Q}^{\mu}\equiv\frac{\delta\mathcal{L}^{\left(4\right)}}{\delta A_{\mu}},
\end{equation}
is set anti-parallel to $\boldsymbol{B}$. In addition, $\epsilon_1$ also impacts the baryon density derived from eq.~\eqref{eq:JB},
specifically
\begin{equation}
J_{B}=-\frac{1}{4\pi^{2}}d\left[\left(f-\sin f\cos f\right)\left(1+\epsilon_1 r\sin\theta A_{\phi}\right)\right]\wedge d\left(\epsilon_2\cos g\right)\wedge d\left(\epsilon_1\phi\right).\label{eq:jbpm}
\end{equation}
Hence the overal sign determined by $\epsilon_1 \cdot\epsilon_2$ represents the particle/antiparticle nature of
the baryon described. It can be shown in a certain circumstance 
that the third homotopy group 
$\pi_3(S^3)$ is a product of 
the winding number of the $\pi^\pm$ phase
along the zeros of $\pi^0$ and  
the winding number of $\pi^0$ phase along the zeros of $\pi^\pm$ 
\cite{Gudnason:2020luj,Gudnason:2020qkd}. 
In our case, the former winding is attached with the $\epsilon_1$ sign 
and the latter with $\epsilon_2$ sign
in the Ansatz (\ref{eq:ansatz}). 
In particular, the ground state Skyrmion and anti-Skyrmion both feature $\epsilon_1=-1$ but the latter has a flipped $\varphi_{3}$: 
\begin{equation}
\bar{\Sigma}\equiv\exp\left(i\boldsymbol{\tau}\cdot\bar{\boldsymbol{\varphi}}\right)=\exp\left\{ if\left[\sin g\left(\cos\phi\tau_{1}+\sin\phi\tau_{2}\right)-\cos g\tau_{3}\right]\right\} .
\end{equation}
So that the baryon number for the anti-Skyrmion
$\bar{\Sigma}$ turns out $N_{{\rm B}}=-1$
seen from eq.~\eqref{eq:jbpm}. For ground
states, $\Sigma$ and $\bar{\Sigma}$ have degenerate masses $M_{-}$
on condition that the winding orientation of $\pi^{\pm}$, i.e., $\epsilon_1$ guarantees
$\boldsymbol{m}\parallel\boldsymbol{B}$. 
The nonmonotonic behavior of $M_-(B)$ is interpreted as the flip of $\boldsymbol{m}$, which was elaborated in ref.~\cite{Chen:2021vou}.
After clarifying such Zeeman-like
physics, for now, we discard considering the solution with $\epsilon_1=+1$, which is for excitation states. 
We remark that
$m_{\text{N}}=M_{\pm}\left(0\right)=864.3\ {\rm MeV}$ is the nucleon mass evaluated in our model.
It bears an $8\%$ discrepancy compared to the
experimental data $938.9\ {\rm MeV}$ since the quantization has not
been taken into account.

\begin{table}
\begin{center}
\begin{tabular}{c|c|c}
             &   $\epsilon_{1}=-1$ (lowest energy) & $\epsilon_{1}=+1$ (excited) \\\hline
        $\epsilon_2=+1$  &  \begin{tabular}{c}
                 $N_{\rm B}=+1$ \\
                $\mathcal{E} = M_- -\mu_{\rm B}$ 
        \end{tabular} 
            &
                \begin{tabular}{c}
                 $N_{\rm B}=-1$\\ 
                 $\mathcal{E} =M_+ + \mu_{\rm B}$
                \end{tabular} \\\hline
  $\epsilon_2=-1$   &      
            \begin{tabular}{c}
                $N_{\rm B}=-1$\\ 
                 $\mathcal{E} =M_- +\mu_{\rm B}$
                \end{tabular}  & 
                \begin{tabular}{c}
                $N_{\rm B}=+1$ \\
                $\mathcal{E} = M_+ -\mu_{\rm B}$
                \end{tabular}
\end{tabular}
\end{center}
\caption{
The baryon numbers
$N_{\rm B}$ 
(topological charges) 
and free energies $\mathcal{E}$ of the configurations solved from Ansatz \eqref{eq:ansatz}.
The Skyrmion mass $M_{\pm}$ is calculated in fig.~\ref{fig:Mpm}.
The diagonal pairs can be transformed to each other by a spatial rotation of the (anti-)Skyrmion by $\pi$ around any $r$-axis that resides in the $xy$ plane. Such a rotation would reverse the direction of  $\boldsymbol{m}$.
\label{table:four-types}}
\end{table}

On top of the (anti-)Skyrmion mass $M$, the contribution of the WZW term to the energy is essentially a chemical potential term in the context of the grand canonical ensemble.
The energy density $\mathcal{E}$ associated with the full Lagrangian $\mathcal{L}^{\left(4\right)}$ is the free energy.  
For $N_{\text{B}}=1$ Skyrmion, it reads $\mathcal{E}\left(\mu_{\rm B},B\right)=M\left(B\right)-\mu_{\rm B}$.
For $N_{\text{B}}=-1$ anti-Skyrmion, it reads $\mathcal{E}\left(\mu_{\rm B},B\right)=M\left(B\right)+\mu_{\rm B}$. Certainly $\mu_{\rm B}$-terms come from $\mathcal{L}_{\text{WZW}}$, an effective way to capture anomaly and density effects in ChPT.
The baryon number and energy of the four types of (anti-)Skyrmions distinguished by $\epsilon_{1,2}$
are summarized in Table~\ref{table:four-types}.
We observe that the true ground state is made of Skyrmions, rather than anti-Skyrmions, as it should be. 

One can immediately observe that when $\mu_{\rm B}$ exceeds the critical value
\begin{equation}
\mu_{c}=M_{-}\left(B\right),
\end{equation}
the free energy of a ground state Skyrmion turns negative, signifying the nucleation of a single Skyrmion from vacuum.  
Such a phenomenon does not occur for anti-Skyrmion, seen from the $+$ sign in front of $\mu_\text{B}$. 
For a Skyrmion, if the $\mu_\text{B}$ is further enhanced to be greater than
\begin{equation}
\mu_{+}=M_{+}\left(B\right)>\mu_c,
\end{equation}
the ground state Skyrmion can be excited, i.e., with a reversed $\boldsymbol{m}$. 

Certainly, the starting point of a phase transition into a Skyrmion crystal shall be the emergence of a ground state Skyrmion with mass $M=M_-$ at $\mu_c(B)$. 
Then if the Skyrmion crystal phase forms, the excited Skyrmion with $M=M_+$ could become relevant because the arrangement of Skyrmions with opposite directions of $\boldsymbol{m}$ in contiguous lattice cells may decrease the crystal energy. The details of the crystalline configuration go beyond the scope of this study. Here, let us first focus on the phase boundary. It is stipulated by $\mu_{c}$ as a function of $B$ shown as the blue curve in the phase diagram fig.~\ref{fig:phase}.
Such a boundary shall be compared with that of CSL, which is given
by eq.~\eqref{eq:muCSL}, and plotted in fig.~\ref{fig:phase} as
the orange curve.
In summary, we mark the phases bounded by different critical $\mu_{B}(B)$ 
curves in the phase diagram fig.~\ref{fig:phase}.

Highlight is the Skyrmion crystal phase at the bottom right corner. It emerges as the 
(large-$N_{\rm C}$) QCD ground state at higher $\mu_{\rm B}$ and lower $B$ compared to the CSL and DWSk phases.
The critical $\mu_{c}(B)$ at $B=0$ is nothing but the nucleon
mass $m_{\text{N}}\equiv\mu_{c}(0)$ predicted by the Skyrme model
as mentioned above. The two curves $\mu_{c}\left(B\right)$
and $\mu_{\text{CSL}}\left(B\right)$ cross at 
\begin{align}
& B^{\ast}=0.368f_{\pi}^{2}s^{2}
\sim 4.55m_{\pi}^{2}, \nonumber \\    
& \mu^{\ast}\equiv\mu_{c}\left(B^{\ast}\right)=\mu_{\text{CSL}}\left(B^{\ast}\right)=72.2f_{\pi}s^{-1}\sim0.932m_{\text{N}}.
\label{eq:Bstar}
\end{align}
This estimation indicates a density window slightly below the nucleon
mass, i.e., $\mu_{\rm B}\in
(\mu^{\ast},\ m_{\text{N}})$, for the nucleon formation at finite $B$. 
Such nucleation could be a 
manifestation of anomaly effects observable in dense matter under
strong magnetic field $B\sim\mathcal{O}(m_{\pi})$.

For $\mu_c<\mu^\ast$, the curve $\mu_c(B)$ is dotted to indicate that it does not delineate the phase boundary between Skyrmion crystal and the vacuum. 
This is because, in that region, the ground state should be the DWSk phase. 
Likewise, the CSL phase boundary 
$\mu_{\rm CSL}(B)$ does not describe the ground state phase transition in $\mu_\text{CSL}> \mu^\ast$ sector (therefore dotted), where the ground state is the Skyrmion crystal.

At density $\mu_\text{B}>\mu^\ast$, it is so far unclear whether there is a phase transition between DWSk and Skyrmion crystal or a crossover. Regarding this issue,  the 
deformation of a Skyrmion cyrstal 
into domain walls found in refs.~\cite{SilvaLobo:2009qw,Leask:2023tti} could be suggestive.
Among the two phases, at higher density and stronger magnetic field, 
the ground state is anticipated to be the DWSk phase, 
because DWSk features not only the topological Skyrmion number but also the term proportional to the magnetic field $B$ in the WZW term, both of which lower the energy. Of course, rigorous further study is needed to make a conclusion. 

Although a Skyrmion crystal could feature lower energy than an isolated Skyrmion~\cite{Chen:2021vou},
the boundary of 
the Skyrmion crystal is set by a sinlge $N_{\text{B}}=1$
Skyrmion, as presented in the current work. The physical procedure
of nuclear matter formation kicks off from one Skyrmion occupying the infinite solution space. 
Then when the chemical potential or density is further enhanced above $\mu_{\rm B}>\mu_{c}$, multiple Skyrmions from afar come closer and assemble to a crystal with finite volume. 
A detailed study on CSL and Skyrmion crystal encompassing finite size effects has been made in ref.~\cite{Chen:2021vou},
although the variation of size is limited to the transverse plane, which means the longitudinal physics highly relevant to $\mu_{\rm B}$ is absent. 
Addressing this line, we outlook the full 3D variation of volume in
order to figure out the phase transition or crossover between Skyrmion crystal and DWSk phase.

Another noteworthy phenonmenon is the baryon crystal proposed in refs..~\cite{Evans:2022hwr,Evans:2023hms} based on the Ginzburg-Landau analysis in the vicinity of density scale $\mu_\text{inst}=4 f_\pi^2 f_\pi^2 / B$ where CSL is meddled by the charged pion condensate
~\cite{Brauner:2016pko}. 
However, the later discovered DWSk phase is supposed to be the ground state at such $\mu_\text{inst}$ 
while CSL is a metastable state. 
Therefore, the conclusion of refs..~\cite{Evans:2022hwr,Evans:2023hms} may need to be reexamined with DWSk taken into account.

\section{Aspects of Skyrmions in a Magnetic Field}
\label{sec:op2}
In this section, we will address several previously underplayed aspects of Skyrmions in a magnetic field. 
The first concerns a conjectured pancake configuration regarding the extrapolation between the Skyrmion and CSL. Our quantification of the Skyrmion scales with the reference of the pancake conjecture would reinforce the understanding of the phase boundary between CSL and Skyrmion crystal. 
The second point is the modified scaling behavior in the presence of a background gauge field, rewriting Derrick's theorem. Such reexamination helps explore the possibility of a Skyrmion without the Skyrme term, pursuing model independent results. 

\subsection{Pancake Conjecture}
As is known, the transverse area of a single soliton in CSL, or equivalently, that of a single DW Skyrmion, is quantized as eq.~\eqref{eq:quantization}, i.e., $S_1=2\pi/eB$. 
If there is indeed a phase transition from Skyrmion crystal into CSL, it is natural to think that near the phase boundary, they share similar configurations. But how can this happen?  
Given the cylindrical symmetry, for a single $\pi^0$ soliton in CSL to configure closer to a Skyrmion, the soliton shall be cut into a pancake shape 
\cite{Amari:2024mip}
whose transverse radius is derived from $S_1$ as  
\begin{equation}
R_\text{P}\equiv\sqrt{\frac{2}{eB}}.
\label{eq:pan}
\end{equation}
Such a quantization condition results purely from the WZW term, whereas there is no dynamical mechanism to prescribe the transverse profile. That is why we call the pancake cut a conjecture.

\begin{figure}      
\centering       
\includegraphics[width=0.75\columnwidth]{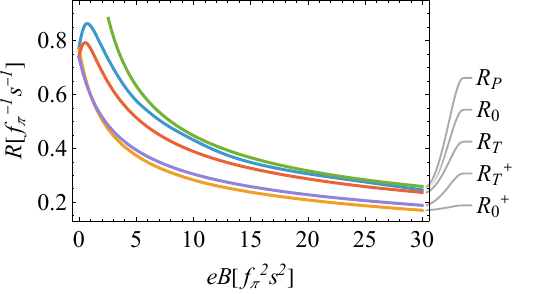}      
\caption{
Characteristic transverse scales of a Skyrmion compared with that of a chiral soliton. $R_\text{P}$ is the radius of the pancake cut of a $\pi^0$ soliton among CSL, namely eq.~\eqref{eq:pan}. $R_0$ defined in eq.~\eqref{eq:R0} is the radius of the superconducting ring where $\pi^\pm$ among $\Sigma$ is maximized on the transverse plane. $R_T$ defined in eq.~\eqref{eq:RT} is the root mean square of the Skyrmion's transverse radius. $R_{0,T}^+$ are the counterparts defined for the excited Skyrmion (the solution branch with soliton mass $M_+$).
}      
\label{fig:R0}  
\end{figure}

On the other hand, the Skyrme model yields an exactly solved Skyrmion profile whose transverse scale is fixed from the dynamics governed by  $\mathcal{L}_{\text{ChPT}}+\mathcal{L}_{\text{Skyr}}$.
There are several characteristic scales, among which we find one that matches $R_\text{P}$ suggestively.
It is a transverse radius $R_0$ defined by the condition 
\begin{equation}
R_0:\quad f\left(R_0,\pi/2 \right)= g\left(R_0,\pi/2\right) =\pi /2,
\label{eq:R0}
\end{equation}
which essentially stipulates the position of a superconducting ring in $xy$ plane where $\pi^\pm$ magnitude $|\sin{f}\sin{g}|$ is maximized.
Here we present a precise quantification of $R_0$ in fig.~\ref{fig:R0}, plotted together with $R_\text{P}$. 

The matching between $R_0$ and $R_\text{P}$ is interpreted as follows. At weak $B$, the phase diagram fig.~\ref{fig:phase} is dominated by nuclear matter where the Skyrmion features a shape deviated from CSL.
In contrast, for higher $B\gtrsim3m_\pi^2$, observed from fig.~\ref{fig:R0}, $R_0$ gets utterly close to $R_\text{P}$. 
The reason is, in this regime, the $B$-related contribution $f_\pi^2\sin^2{f}\sin^2{g}\Upsilon_{\pm}^2/2$ to the energy density \eqref{eq:T00} is dominant. 
To minimize such a portion of energy, the $\pi^\pm$ is highly localized to the ring with transverse radius 
\begin{equation}
R_0(B\rightarrow\infty)\rightarrow\mathrm{argmin}\Upsilon^2_{\pm}\left(r\mathrm{sin}\theta\right) = R_\text{P},  
\label{eq:R0P}
\end{equation} 
which coincides with $R_\text{P}$ defined by eq.~\eqref{eq:pan}.
Such localization at strong $B$ is essential to explain the extrapolation from a Skyrmion to a pancake $\pi^0$ soliton. 
Therein the Skyrmion configuration highly deformed by $B$ is almost occupied by $\pi^0$ everywhere except at the ring with $r\, \mathrm{sin}\theta=R_0$. Inside the ring $r\, \mathrm{sin}\theta<R_0$ the $\pi^0$ winds $2\pi$ between $z=\pm \infty$, same with its behavior in CSL.
Outside the ring $r\, \mathrm{sin}\theta > R_0$ the $\pi^0$ remains almost homogeneous vacuum as it does on the spatial boundary $\Sigma(r\rightarrow\infty)=\mathbb{I}$.
Hence the only difference between such a magnetically deformed Skyrmion and the pancake $\pi^0$ soliton lies in that the thin $\pi^\pm$ ring provides a distinguished homotopy but minimal kinetic energy contribution. 

For a double check of such an interpretation, in addition to $R_0$, we try to capture the transverse size of the Skyrmion.  One way is evaluating the mean square transverse radius weighted by the baryon density as a distribution function.
In general, one can exploit either the topological baryon density
\begin{equation}
n_\text{B}=-\frac{\epsilon^{0\alpha\beta\gamma}}{4\pi^2}l_\alpha l_\beta l_\gamma,
\end{equation}
or the covariant $J_\text{B}^0$ derived from eq.~\eqref{eq:JB}. In our setup, the latter is less proper since $B$ remains finite on the spatial boundary at infinity (rather than a pure gauge), leading to possible divergence of physical quantities evaluated with $J_\text{B}^0$ at large $B$, which is an artifact that could be reconciled by promoting the gauge field to be dynamical.
Meanwhile, $n_\text{B}$ is qualified to represent the density distribution of the Skyrmion since the normalization $\int{n_\text{B}d^3x}=\int{J_\text{B}^0 d^3x}$ proves valid regardless of $B$.
Henceforth, we specify the definition of transverse radius in the present work as
\begin{equation}
R_{T}^2\equiv \int d^3x n_\text{B} \left(r \sin{\theta}\right)^2.
\label{eq:RT}
\end{equation}
We would apply $n_\text{B}$ to evaluate the average of other physical quantities as well, which will soon be seen in the next subsection.
The resulting $R_T(B)$ is plotted in fig.~\ref{fig:R0}. 
It tends to converge with $R_0$ at large $B$, which reinforces our intuition that in transverse dimensions, the deformed Skyrmion is almost bounded by the $\pi^\pm$ superconducting ring. 
In other words, the Skyrmion's configuration outside the ring is nearly the unified vacuum.
Furthermore, the closeness between $R_{0,T}$ and $R_\text{P}$ then demonstrates a Skyrmion deformed by a large $B$ resembles a $\pi^0$ soliton (winding in CSL style) cut to a pancake shape, except that the Skyrmion is bounded by a thin $\pi^\pm$ ring.
We argue the ring is thin because to minimize the energy which is dominated by the contribution from $\Upsilon_\pm$-term at large $B$, the $\pi^\pm$ is highly localized into the position with $r\sin{\theta}=R_0$ where $\Upsilon_\pm^2$ takes the minimal value, as can be seen from eq.~\eqref{eq:upsilon}. 
In this way, we have explained the two configurations and their relation at large $B$. 
Certainly, our description is heuristic for an intuitive understanding of the conjectured phase transition. 
A more solid physical argument relies on a full 3D analysis of a Skyrmion crystal with finite size effects included.  

For curiosity, in fig.~\ref{fig:R0}, we added $R^+_{0}$ and $R^+_T$, which are the counterparts of $R_{0}$ and $R_T$ 
derived from the $f$ and $g$ of the excited Skyrmion solution with soliton mass $M_+$. 
The smallness of $R^+_{0,T}$ compared to $R_{\text{P},0,T}$ hints that the excited Skyrmion is of higher density and energy, as it should be. 
Thus, we reiterate the solution branch with $M_+$ is irrelevant to our discussion on the phase boundary that involves only ground states.  
No matter being an excited state or not, at the full range of $B$, Skyrmion bears a mean transverse size smaller than that of a pancake $\pi^0$ soliton, e.g., seen from $R_{0,T}<R_\text{P}$, which is consistent with the fact that the Skyrmion crystal conquers the ground state at higher density $\mu_\text{B}$ than CSL, as also seen from our central result Fig,~\ref{fig:phase}.

Apart from the implication on the phase structure, some general aspects of the Skyrmion property are remarked at last.
The nonmonotonic behavior of $R_{0,T}(B)$ is consistent with that of $M(B)$ as seen from fig.~\ref{fig:Mpm}, the mechanism of which lies in Zeeman-like physics detailed in Sec.~\ref{sec:SkX} as well as ref.~\cite{Chen:2021vou}.
In short, the competition between the $R_{P}$ scale controlled by $B$ and the inherent size of a Skyrmion, i.e., nucleon size without effects of $B$, leads to the summit of $R_{0,T}(B)$.

\subsection{Without Skyrme Term?}
The motivation to discuss a Skyrmion without Skyrme term is to see if the above result could be model-independent, i.e., independent
of $s$, from pure ChPT at strong $B$. In theory, the possibility
of a stable $\mathcal{O}\left(p^{2}\right)$ gauged Skyrmion solution
is not ruled out by the scaling law, 
as explicated in Appendix \ref{sec:Derrick}. 
In fact, in 2+1 dimensions, gauged O(3) lumps are unstable without a potential 
but can be stabilized by a background magnetic field \cite{Amari:2024adu}.
Another example can be found in DW Skyrmions 
\cite{Amari:2024fbo}. 
Specifically, DW Skyrmions are stable for large gauge coupling, while 
DW anti-Skyrmions are stable 
in the whole parameter range 
without a Skyrme term. 
Such altered scaling behavior is exclusive
to the case of a background gauge field whose scaling transformation
is not entirely trivial as thought.

\begin{figure}[!htb] 
\minipage{0.5\textwidth}   
\includegraphics[width=\linewidth]{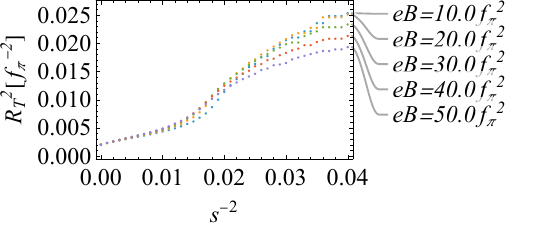}   
\endminipage\hfill 
\minipage{0.5\textwidth}   
\includegraphics[width=\linewidth]{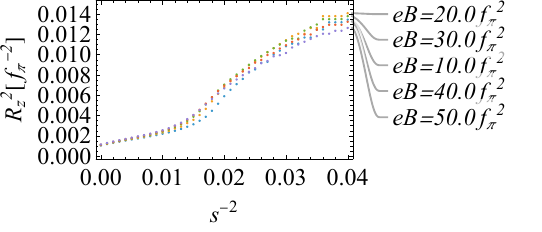}   
\endminipage\hfill 
\caption{Mean square radii at fixed values of $B$ with a changing $s$.}
\label{fig:radii}
\end{figure}

However our numerical efforts searching for the $\mathcal{O}\left(p^{2}\right)$ Skyrmion at finite $B$ turn out
frustrating results. We exploit dimensionless quantities via $x^{\mu}\rightarrow\tilde{x}^{\mu}=xf_{\pi}$
and then diminish $s^{-2}$ from the default input $4.84^{-2}$ to $0$,
aiming at a Skyrmion profile with finite $B$ and vanishing Skyrme
term. Our observation is as follows: the profile itself is conformal
under varying $s$, yet with the size shrinking with a descending
$s^{-2}$, which eventually leads to the collapse of the Skyrmion
towards null when $s^{-2}\rightarrow0$. 
We demonstrate this
point quantitatively via fig.~\ref{fig:radii}, the plot of typical scales of Skyrmion size, the mean square of transverse radius \eqref{eq:RT} and longitudinal radius defined by
\begin{equation}
R_{z}^2\equiv \int d^3x n_{\text{B}} \left(r \cos{\theta}\right)^2.
\end{equation}

\begin{figure}      
\centering       
\includegraphics[width=0.75\columnwidth]{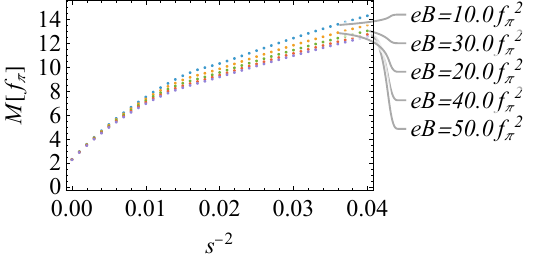}      
\caption{Skyrmion mass at fixed values of $B$ with a changing $s$.}     
\label{fig:M}  
\end{figure}

One may find that around $s^{-2}\rightarrow 0$, $R_{z,T}$ does not vanish completely. We claim this is due to the limitation of numerical precision in the present finite element approach at small system size, which is a technical artifact to be overcome. To further convince that the gauged Skyrmion fails to survive $s^{-2}\rightarrow0$ limit, we present the Skyrmion mass $M$ as a functional of $s^{-2}$ and $B$ in fig.~\ref{fig:M}. It shows $M\left( s^{-2}\rightarrow0, B\neq0\right)\rightarrow 0$ albeit numerical artifact again.
Conclusively, $\mathcal{O}(p^{2})$ Skyrmions can not be
stabilized by a background magnetic field.

\section{Summary and Discussion}
\label{sec:summary}

In this paper, we have determined the phase boundary between 
the Skyrmion crystal and QCD vacuum 
in the QCD phase diagram  in the 
background magnetic field $B$ 
at finite baryon chemical potential $\mu_{\rm B}$. 
Starting from $\mu_c =m_N$ the nucleon mass at $B=0$, 
the phase boundary bends to smaller $\mu_{\rm B}$ for larger $B$, 
and finally ends on the CSL boundary with intersection $(\mu^\ast\sim 0.932 m_\text{N},\,B^\ast\sim 4.55 m_\pi^2)$, as shown in fig.~\ref{fig:phase} and eq.~\eqref{eq:Bstar}.

We have further examined the Son and Stephanov's conjecture that 
a single Skyrmion could emerge as a pancake shaped $\pi^0$ domain wall winding inside a $\pi^\pm$ superconducting ring.
Our exact numerical results support the pancake imitation of Skyrmion profile at large $B$ with the fact that Skyrmion transverse radius in eq.~\eqref{eq:RT} and superconducting ring radius in eq.~\eqref{eq:R0P} tend to converge, nearly satisfying the quantization condition~\eqref{eq:quantization} with $S_{N_{\rm B}}/N_{\rm B}=\pi R_P^2\sim \pi R_0^2 \sim \pi R_T^2$.
Meanwhile, at smaller $B$ the Skyrmion configuration deviates from the pancake, as one can see from fig.~\ref{fig:R0}, in which case the Skyrmion shall be described as a prolate ellipsoid.

In addition, we have numerically manifested that, even if in the background magnetic field, Skyrmions are unstable at ${\cal O}(p^2)$ (the profile would shrink to vanish, c.f. fig.~\ref{fig:radii}) although the Derrick's scaling argument does not rule out the possibility of a stable solution in theory, as detailed in Appendix \ref{sec:Derrick}.

Albeit we have explored the phase boundary involving CSL and the Skyrmion crystal, pinning down the Skyrmion crystal structure for dense nuclear matter in a magnetic field remains a future task. 
Only through such study can we simulate rigorously the transition between multiple low energy topological phases in the baryon rich context. 
Moreoever, on the phase boundary, the phase transition belongs to the so-called nucleation type. 
The quantum~\cite{Eto:2022lhu,Higaki:2022gnw} 
and dynamical~\cite{Eto:2025ebz} nucleation of CSL was studied. Similar demonstration should be further developed for 
the creation of Skyrmions in the bulk.

Finally, we comment on the chiral limit $m_\pi\to 0$ for discussions.
Skyrmion solutions are not significantly modified by changing the input $m_\pi$ to zero, as known from literature such as~\cite{Adkins:1983ya,Adkins:1983hy} for $B=0$ case and~\cite{Chen:2021vou,Chen:2023jbq} incorporating $B$. 
Hence, the blue curve 
in the phase diagram in fig.~\ref{fig:phase} 
would remain nearly unchanged in chiral limit. 
However, in contrast, the critical magnetic field of CSL
in eq.~(\ref{eq:quantization})
or equivalently the critical chemical potential $\mu_\text{CSL}$ in Eq.~(\ref{eq:muCSL}),
approaches zero, indicating that CSL could arise immediately upon turning on the magnetic field, no matter how small it is. 
In view of the phase diagram, the orange curve in fig.~\ref {fig:phase} would approach the two axes, above which the CSL phase is prone to occupy the full quadrant. In other words, the uniform vacuum is almost occupied by CSL.
Then the entire blue curve in fig.~\ref{fig:phase} becomes dotted, and the Skyrmion crystal phase resides within the CSL phase. 
We can not determine such a phase boundary from the single Skyrmion analysis presented in this paper. 
Meanwhile, the phase boundary between the DWSk and CSL phases, i.e., the 
green curve in fig.~\ref{fig:phase}, tends to a vertical line with $\mu_\text{DWSk}\rightarrow\infty$ seen from Eq.~\eqref{eq:muDW}, which ends at the point $(\mu_D,B_D) \to (\infty,0)$. 
That means the boundary goes deep inside the Skyrmion crystal phase.   
Therefore, the chiral limit requires delicate analysis on the following issues: Which of the Skyrmion crystal and CSL is energetically favored? Is the phase transition crossover? Under what condition can the transition occur? There are things to find out.

\section*{Acknowledgment}
This work is supported in part by JSPS KAKENHI [Grants 
No.~JP23KJ1881 (Y.~A.)
No.~JP22H01221 and 
No.~JP23K22492 
(M.~N. and Z.~Q.)],
and 
the WPI program ``Sustainability with Knotted Chiral Meta Matter (WPI-SKCM$^2$)'' at Hiroshima University (M.~N.).

\begin{appendix}

\section{Derrick's Theorem with a Background Gauge Field}\label{sec:Derrick}
In this appendix, we show that Derrick's scaling argument \cite{Derrick:1964ww} does not rule out the possibility of stable soliton solutions in the ${\cal O}(p^2)$ theory, though the numerical analysis in Sec. III suggests that such solutions do not exist.
The static energy associated with the Lagrangian \eqref{eq:Lkin}
is given by
\begin{align}
    E_{\text{kin}}&=\int d^3 x \left[ 
     -\frac{f^2_\pi}{4}\Tr\left(L_iL_i \right) 
     +\dfrac{f_\pi^2m_\pi^2}{2}\Tr(\mathbf{1}-\Sigma)
     \right]
\end{align}
For convenience of notation, we decompose the energy as
\begin{equation}
    E_{\text{kin}}=E_2^{(r)} + E_2^{(z)} + E_1 + E_0^{A} + E_0^{_{\text{mass}}}
\end{equation}
with 
\begin{align}
    &E_2^{(r)} = -\frac{f^2_\pi}{4}\int d^3 x
    \Tr\left(\Sigma^\dagger \partial_k \Sigma \Sigma^\dagger \partial_k \Sigma \right) 
    \ ,
    \\
    &E_2^{(z)} = -\frac{f^2_\pi}{4}\int d^3 x
    \Tr\left(\Sigma^\dagger \partial_z \Sigma \Sigma^\dagger \partial_z \Sigma \right) 
    \\
    &E_1 = -\frac{f^2_\pi}{4}\int d^3 x ~ 2ie
     A_k\Tr\left( \Sigma^\dagger \partial_k \Sigma\Sigma^\dagger[Q,\Sigma]\right) 
    \ , \\
    &E_0^A = \frac{f^2_\pi}{4}\int d^3x  ~ e^2A_k^2 \Tr\left( \Sigma^\dagger[Q,\Sigma] \Sigma^\dagger[Q,\Sigma]\right) \ ,
    \\
   & E_0^{_{\text{mass}}}=\dfrac{f_\pi^2m_\pi^2}{2}\int d^3 x       
     \Tr(\mathbf{1}-\Sigma) \ .
\end{align}
where $k=1,2$ and the indices represent the number of the partial derivative.
Suppose that $\Sigma(x,y,z)$ describes a localized static solution. 
Since the background magnetic field is applied along the $z$-axis, 
scaling property in the longitudinal and transverse directions would be different.
Thus, we consider two types of transformation
\begin{align}
    &\Sigma(x,y,z) \to \Sigma_\lambda(x,y,z) =\Sigma(\lambda x, \lambda y, z)
    \label{eq:scaling_xy}
    \\
    &\Sigma(x,y,z) \to \Sigma_\xi(x,y,z) = \Sigma(x,  y, \xi z)
    \label{eq:scaling_z}
\end{align}
with a positive constant $\lambda$ and $\xi$.
We shall derive the virial relations associated with each scaling \eqref{eq:scaling_xy} and \eqref{eq:scaling_z}, which are conditions that finite energy static soliton solutions must satisfy.
Note that those give a more strict condition than the virial relation obtained through the usual homogeneous scaling $\Sigma(x,y,z) \to \Sigma(\zeta x, \zeta y, \zeta z)$.
Since the gauge potentials are background fields, they do not change according to these transformations.

Firstly, we consider the transverse scaling \eqref{eq:scaling_xy}. 
Noting that the gauge potential satisfies
\begin{align}
     A_k(x,y,z) = \lambda^{-1} A_k(\lambda x, \lambda y, z) \ ,
\end{align}
one can write the energy for $\Sigma_\lambda$ as
\begin{align}
    &e^{(r)}(\lambda)\equiv 
    E_{\text{kin}}[\Sigma_\lambda]
    \notag\\
    &= -\frac{f^2_\pi}{4}\int d^3 x\left[
     \Tr\left(\Sigma^\dagger_\lambda \partial_a \Sigma_\lambda  \right)^2+\Tr\left(\Sigma^\dagger_\lambda \partial_z \Sigma_\lambda  \right)^2
     \right.
     \notag\\
     &\left. \qquad\qquad\qquad\qquad
     -2ieA_k\Tr\left(  \partial_a\Sigma^\dagger_\lambda[Q,\Sigma_\lambda]\right)
     -e^2A_k^2 \Tr\left( \Sigma^\dagger_\lambda[Q,\Sigma_\lambda] 
      \right)^2
      -2m_\pi^2\Tr(\mathbf{1}-\Sigma_\lambda)
    \right] 
    \notag\\
    &= -\frac{f^2_\pi}{4}\int d^3 x\left[
     \lambda^2\Tr\left(\Sigma^\dagger_\lambda 
     \frac{\partial \Sigma_\lambda}{\partial (\lambda x^k)}  \right)^2+\Tr\left(\Sigma^\dagger_\lambda \partial_z \Sigma_\lambda  \right)^2
    \right.
    \notag \\
    &\left. \qquad\qquad\qquad\qquad
     -2\lambda ieA_k({\bm x})
    \Tr\left(  \frac{\partial \Sigma^\dagger_\lambda}{\partial (\lambda x^k)}[Q,\Sigma_\lambda]\right)
     -e^2A_k^2({\bm x}) 
     \Tr\left( \Sigma^\dagger_\lambda[Q,\Sigma_\lambda] \right)^2
     -2m_\pi^2\Tr(\mathbf{1}-\Sigma_\lambda)
    \right] 
    \notag\\
    &= -\frac{f^2_\pi}{4}\int d (\lambda x) d(\lambda y) dz\left[
     \Tr\left(\Sigma^\dagger_\lambda 
     \frac{\partial \Sigma_\lambda}{\partial (\lambda x^k)}  \right)^2
     +\lambda^{-2}\Tr\left(\Sigma^\dagger_\lambda \partial_z \Sigma_\lambda  \right)^2
    \right.
    \notag \\
    & \qquad\qquad\qquad\qquad\qquad\qquad
     -2\lambda^{-2} ieA_k({\lambda x, \lambda y,z })
    \Tr \left(  \frac{\partial \Sigma^\dagger_\lambda}{\partial (\lambda x^k)}[Q,\Sigma_\lambda]\right)
    \notag\\
    &\left. \qquad\qquad\qquad\qquad\qquad\qquad
     -\lambda^{-4}e^2A_k^2({\lambda x, \lambda y,z })
     \Tr\left( \Sigma^\dagger_\lambda[Q,\Sigma_\lambda] \right)^2
     -2\lambda^{-2}m_\pi^2\Tr(\mathbf{1}-\Sigma_\lambda)
    \right] 
    \notag \\
    &=E_2^{(r)} + \lambda^{-2} \left( E_2^{(z)} +E_1 +E_0^{\text{mass}}\right) + \lambda^{-4} E_0^A \ .
\end{align}
If $\Sigma(\bm{x})$ is a static solution, $e^{(r)}(\lambda)$ should be stationary at $\lambda=1$ and therefore we get the virial relation
\begin{equation}
    \left. \frac{de^{(r)}(\lambda)}{d\lambda}\right|_{\lambda=1} = -2 (E_2^{(z)}+E_1+E_0^{\text{mass}}) -4 E_0^A =0
    \quad
    \Rightarrow \quad E_2^{(z)} + E_1 + E_0^{\text{mass}}+ 2E_0^A =0 \ .
    \label{eq:virial_xy}
\end{equation}
Since $E_1$ is not positive semi-definite, the relation is possibly satisfied.

Next, we consider the transformation \eqref{eq:scaling_z}. In this case, the gauge potential satisfy
\begin{align}
    & A_k(x,y,z) =A_k(x,  y, \xi z) \ .
\end{align}
Therefore, the energy for $\Sigma_\xi$ can be written as
\begin{align}
    &e^{(z)}(\xi)\equiv 
    E[\Sigma_\xi]
    \notag\\
    &=-\frac{f^2_\pi}{4}\int d x dy d(\xi z)\left[
     \xi^{-1}\Tr\left(\Sigma^\dagger_\xi 
     \partial_a \Sigma_\xi \right)^2
     +\xi\Tr\left(\Sigma^\dagger_\xi \frac{\partial \Sigma_\xi}{\partial (\xi z)} \right)^2
    \right.
    \notag \\
    &\qquad\qquad\qquad\qquad\qquad\quad
     -2\xi^{-1} ieA_k({ x, y,\xi z })
    \Tr \left(  \partial_a \Sigma^\dagger_\xi[Q,\Sigma_\xi]\right)
    \notag \\
    &\left. \qquad\qquad\qquad\qquad\qquad\quad
     -\xi^{-1}e^2A_k^2({x,  y, \xi z })
     \Tr\left( \Sigma^\dagger_\xi[Q,\Sigma_\xi] \right)^2
     -2\xi^{-1}m_\pi^2\Tr(\mathbf{1}-\Sigma_\xi)
    \right] 
    \notag \\
    &=\xi E_2^{(z)} + \xi^{-1} \left( E_2^{(r)} +E_1 + E_0\right) \ .
\end{align}
So, we obtain the virial relation associated with the scaling \eqref{eq:scaling_z} as
\begin{equation}
    \left. \frac{de^{(z)}(\xi)}{d\xi}\right|_{\xi=1} = E_2^{(z)} - (E_2^{(r)}+E_1 + E_0^A + E_0^{\text{mass}}) =0
    \quad
    \Rightarrow \quad E_2^{(z)} = E_2^{(r)}+E_1 + E_0^A + E_0^{\text{mass}}   \ .
    \label{eq:virial_z}
\end{equation}
This condition can also be satisfied. 

The two virial relations \eqref{eq:virial_xy} and \eqref{eq:virial_z} seem not to be contradictory. Therefore, the theory successfully evades Derrick's no-go theorem and is not ruled out as possessing static finite energy soliton solutions in the scaling argument. However, the numerical results deny the existence of such solutions.

\end{appendix}

\bibliographystyle{apsrev4-1}
\bibliography{references}

\end{document}